\documentclass[journal=jctcce]{achemso}
\setkeys{acs}{articletitle = true}
\usepackage{amsfonts,amssymb,amsmath,latexsym}
\usepackage[english]{babel}
\usepackage{xcolor}
\usepackage{setspace}
\usepackage{bm}
\usepackage{dcolumn}
\usepackage{graphicx}
\usepackage{tabularx}
\usepackage{float}
\usepackage{bigstrut}
\usepackage{subcaption}

\usepackage[normalem]{ulem}

\newcommand{\lcpq}{Laboratoire de Chimie et Physique Quantiques, CNRS,
Universit\'e de Toulouse 3 Paul Sabatier,
118 route de Narbonne, Toulouse, 31062 France}
\newcommand{\warwick}{Department of Physics, University of Warwick, Coventry CV4 7AL, United Kingdom}
\newcommand{\etsf}{European Theoretical Spectroscopy Facility (ETSF)}
\newcommand{\unibo}{Universit\`a di Bologna,
Via Irnerio 33, 40126 Bologna, Italy}

\title{ 
Solution to the Thomson problem for Clifford tori with an application to Wigner crystals
}

\author{Amer Alrakik}
\affiliation{\lcpq}
\author{Miguel Escobar Azor}
\affiliation{\lcpq}
\alsoaffiliation{\warwick}
\alsoaffiliation{\etsf}
\author{V\'eronique Brumas}
\affiliation{\lcpq}
\author{Gian Luigi Bendazzoli}
\affiliation{\unibo}
\author{Stefano Evangelisti}
\email{stefano.evangelisti@irsamc.ups-tlse.fr}
\affiliation{\lcpq}
\author{J. Arjan Berger}
\email{arjan.berger@irsamc.ups-tlse.fr}
\affiliation{\lcpq}
\alsoaffiliation{\etsf}
\date{\today}

\begin{document}

\begin{tocentry}
\includegraphics{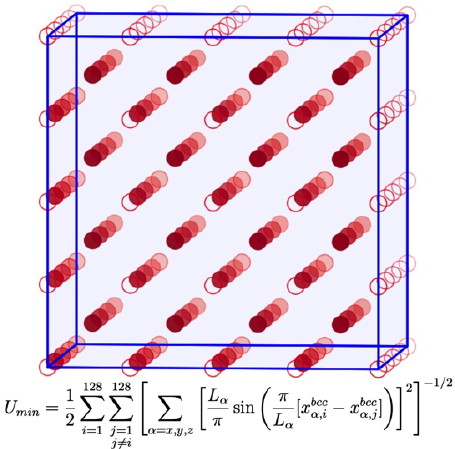}
\end{tocentry}

\begin{abstract}

In its original version, the Thomson problem consists of the search for the minimum-energy
configuration of a set of point-like electrons that are confined to the surface of a two-dimensional 
sphere (${\cal S}^2$) that repel each other according to Coulomb's law, in which the
distance is the Euclidean distance in the embedding space of the sphere, {\em i.e.}, $\mathbb{R}^3$. 
In this work, we consider the analogous problem where the electrons are confined to an 
$n$-dimensional flat Clifford torus ${\cal T}^n$ with $n = 1, 2, 3$. 
Since the torus ${\cal T}^n$ can be embedded in the complex manifold $\mathbb{C}^n$, 
we define the distance in the Coulomb law as the Euclidean distance in $\mathbb{C}^n$, 
in analogy to what is done for the Thomson problem on the sphere. 
The Thomson problem on a Clifford torus is of interest because super-cells with the topology of Clifford torus can be used to describe periodic systems such as Wigner crystals.
In this work we numerically solve the Thomson problem on a square Clifford torus.
To illustrate the usefulness of our approach we apply it to Wigner crystals.
We demonstrate that the equilibrium configurations we obtain for a large numbers of electrons are consistent with the predicted structures of Wigner crystals.
Finally, in the one-dimensional case we analytically obtain the energy spectrum and the phonon dispersion law. 
\end{abstract}

\maketitle

\section{Introduction}

In 1904, Joseph John Thomson, the discoverer of the electron, proposed a model for the atomic structure that later became known as the plum-pudding model~\cite{thomson1904xxiv}. 
In this scheme, the electrons are treated as point-like charges embedded in a large, positively charged spherical nucleus.
This description soon turned out to be incorrect and was quickly abandoned and replaced by the Rutherford's solar-system model.
Today, the plum-pudding model has essentially just a historical interest.
This approach, however, gave rise to a very prolific mathematical literature, centered on the so-called Thomson problem.
This problem aims to find the minimum-energy configuration of $N$ point-like electrons that are confined to the surface of a two-dimensional sphere and subject to the repulsive force described by Coulomb's law.
For a small number of electrons the exact solution to this problem is sometimes known, even though in some cases the proof is far from trivial~\cite{schwartz2013five}.
Numerical solutions, instead, are available up to very large values of $N$ (several hundreds).
We note that the original Thomson problem has been generalized to a wide class of potentials and to a various number of spatial dimensions~\cite{Rakhmanov_1994,Kuijlaars_1998,Martinez_2004,Cioslowski_2008_PRE,Cioslowski_2009_PRE,Cioslowski_2009_JCP,Cioslowski_2010_JCP,Cioslowski_2011_JCP,Cioslowski_2012_JCP,Cioslowski_2012_JMC,Cioslowski_2014_JMC,Cioslowski_2013_JCP,Cioslowski_Albin_2013_JCP}.

In chemistry, the Thomson problem has found a direct application in the valence-shell electron-pair repulsion (VSEPR) theory of Gillespie and Nyholm, which permits to predict the local shape of a molecule by minimizing the Coulomb repulsion energy of negatively charged electron pairs that surround a positively charged nucleus\cite{gillespie1957inorganic}.
The VSEPR formalism concerns a small number of electron pairs only, from 2 pairs, \emph{e.g.} BeH$_2$, up to at most 9 pairs,  \emph{e.g.} ReH$_9^{2-}$ in the K$_2$ReH$_9$ ionic crystal.
In physics and chemistry, applications of the Thomson problem to a large numbers of electrons have been relatively rare up to now. 

In a series of recent papers~\cite{tavernier2020clifford,tavernier2021clifford,alves2021accurate,escobar2021wigner} 
we have shown that periodic, $n$-dimensional crystalline systems can be conveniently mapped onto a Clifford torus ${\cal T}^n$. 
One of the reasons for the success of this approach is that Clifford tori are flat manifolds, which means that a fragment of a crystal can be modified to the topology of a Clifford torus without deformation.
We note that, on the contrary, a similar statement is, in general, not true for spheres, \emph{i.e.}, a fragment of a crystal \emph{cannot} be modified to the topology of a sphere without deformation.
For this reason, mapping super-cells onto spheres is not a particularly useful model for two- and three-dimensional crystals.
An interesting application of our approach is the mapping of Wigner crystals on a Clifford torus.
A Wigner crystal is a crystalline structure that forms in a system of interacting electrons at sufficiently low density, in which the electrons localize at periodic lattice sites. This concept was first proposed by Eugene Wigner in 1934, who predicted that in a neutralizing uniform background, the repulsive interactions between electrons would dominate over the kinetic energy of the electrons at low densities.~\cite{wigner1934interaction,wigner1938effects} This would result in the formation of a crystalline structure where the electrons occupy the lattice sites, leading to a decrease in the potential energy of the system. 
A two-dimensional Wigner crystal has recently been observed experimentally by Smolenski and coworkers~\cite{Smolenski_2021}.
Two-dimensional crystal lattices made up of electrons have also be obtained experimentally on the surface of liquid Helium~\cite{Grimes_1979_2D}, in magnetic fields~\cite{Lozovik_1975,Andrei_1988,Goldman_1990,Williams_1991,Ye_2002, Chen_2006, Jang_2017} and in Moir\'e superlattices~\cite{Regan_2020,Tang_2020,Xu_2020,Li_2021}.
Such systems are often also referred to as Wigner crystals in the literature.
Wigner crystals have also been observed in one dimension~\cite{Deshpande_2008,Shapir_2019,Ziani_2021}.
The properties of Wigner crystals have been studied extensively in condensed matter physics and have important implications for the understanding of the behavior of electrons in low-dimensional systems.
Closely related to Wigner crystals are Wigner molecules, which are confined few-electron systems in which the electrons form a stable bound state due to their mutual repulsion
~\cite{Cioslowski_2006,Ellenberger_2006,Yannouleas_2007,Mendl_2014,Cioslowski_2017,Cioslowski_2017_JCP,Egger,Diaz-Marquez_2018,telleria2022wigner,escobar2019wigner,escobar2021wigner}.
Experimental observations of Wigner molecules have been reported in various physical systems, including carbon nanotubes~\cite{Pecker2013}, nanowires~\cite{Mendez-Camacho2022_1,Mendez-Camacho2022_2} and in quantum dots~\cite{Thakur_2022}.

In our previous works, we have shown that a set of point-like electrons placed on Clifford tori ${\cal T}^n$ of different dimensions ($n$=1,2,3) can efficiently represent a fragment of a classical Wigner crystal~\cite{escobar2019wigner,alves2021accurate,escobar2021wigner}.
In particular, we have been able to compute the classical electrostatic energies as well as the harmonic corrections of Wigner crystals with unprecedented precision~\cite{alves2021accurate}.
This precision has been attained thanks to an extrapolation of the results obtained for finite systems.
An advantage of calculations performed using Clifford periodic boundary conditions, and contrary to calculations performed within open-boundary conditions, is that there are no border effects which  hamper the extrapolation of the results obtained for finite systems to the thermodynamic limit.
One of the main issues in the study of Wigner crystals remains the prediction of the energy and stability of the different possible space groups that the system can adopt and, in particular, the prediction of the space group that corresponds to the ground state.
Usually the ground-state configuration is obtained by comparing the energies of only a few well-known space groups.
However, this leaves open the possibility that a less trivial configuration has a lower energy.
In this paper, we tackle this problem by generalizing the Thomson problem to Clifford tori.
We will show that this approach permits us to predict the geometrical structure of Wigner crystals.

This article is organized as follows.
In section \ref{Sec:Thomson}, we give the details of the Thomson problem on Clifford tori.
We give the computational details of our numerical approach in section \ref{Sec:CompDet}.
In section \ref{Sec:Results}, we report and discuss some results we obtain with our approach.
We will focus on Clifford tori that contain numbers of electrons that are compatible with Wigner crystals.
We discuss the analytical treatment of the one-dimensional case in section \ref{Clifford1D}.
Finally, in Section \ref{Sec:Conclusions} we draw the conclusions of our work and briefly discuss how our approach can be generalized to perform periodic molecular mechanics simulations.
\section{The Thomson Problem on a Clifford torus}
\label{Sec:Thomson}
The original Thomson problem pertains to point-like electrons that are confined to the surface of a two-dimensional sphere and that repel each other according to Coulomb's law and the goal is to find the configuration of the electrons that corresponds to the lowest energy.
The distance in the Coulomb interaction is defined as the Euclidean distance in the embedding space of the sphere, {\em i.e.}, $\mathbb{R}^3$. 
Here we consider the analogous problem where the electrons are confined to an $n$-dimensional (nD) Clifford torus ${\cal T}^n$ with $n = 1, 2, 3$.
A Clifford torus is obtained by joining the opposite points of a line (1D), the opposite edges of a rectangle (2D), or the opposite faces of a cuboid (3D), without deformation.
Analogous to the original problem, we define the distance in the Coulomb potential as the Euclidean distance in the embedding space of the Clifford torus, {\em i.e.}, $\mathbb{C}^n$.
This distance $d_{ij}$ between electrons $i$ and $j$ is thus defined as~\cite{Aragao_2019,tavernier2020clifford,tavernier2021clifford,alves2021accurate,escobar2021wigner}
\begin{equation} 
    \label{euclidian_distance}
    d_{ij} = \sqrt{ \sum_{\alpha=1}^n \left[\frac{L_{\alpha}}{\pi} \sin  \left(\frac{\pi}{L_{\alpha}}[{x_{\alpha,i}}-{x_{\alpha,j}}]\right)\right]^2},
\end{equation} 
where $L_{\alpha}$ is the length of an edge along the $\alpha$-direction of the Clifford torus.
The Coulomb potential $U$ is given by (in Hartree atomic units)
\begin{equation}
U = \frac12\sum_{i=1}^N\sum_{\substack{j=1 \\ j\neq i}}^N d_{ij}^{-1},
\label{Eqn:totenergy}
\end{equation}
with $d_{ij}$ given in Eq.~\eqref{euclidian_distance}.
The Thomson problem on a Clifford torus consists of minimizing the corresponding Coulomb energy with respect to the positions of the electrons on the Clifford torus.
In order to guarantee that a minimum is obtained the gradient of $U$ has to vanish and all the eigenvalues of the Hessian of $U$ should be non-negative while exactly $n$ eigenvalues should be equal to zero. They correspond to the $n$ translational degrees of freedom in the Clifford torus.
Furthermore, there are $n(n-1)/2$ eigenvalues that asymptotically vanish when $L$ tends to infinity for a fixed number of electrons.
They correspond to rotational degrees of freedom.
For example, in the 3D case, there are three eigenvalues that are identically zero corresponding to the translations, and three asymptotically vanishing eigenvalues corresponding to the rotations.
Finally, we note that the Thomson problem in one dimension is equivalent to the well-known Thomson problem on a circle of radius $R$ and length $L = 2\pi R$.
The minimum-energy configuration on the circle is obtained by placing the $N$ electrons on the vertices of a regular polygon with $n$ sides.
On the 1D Clifford torus this corresponds to the positions of the electrons being equidistant.
A complete analytical treatment of the 1D case is presented in section \ref{Clifford1D}.

In two and three dimensions we can, in principle, solve the Thomson problem numerically for any number of electrons with our approach.
For a large number of electrons we expect that the positions of the electrons correspond to the lattices of Wigner crystals, i.e., a hexagonal (or triangular) lattice in two dimensions and a body-centred cubic lattice in three dimensions~\cite{Sholl_1967,hasse1991structure,alves2021accurate}.
\section{Computational Details}
\label{Sec:CompDet}
We wrote a computer code to numerically solve the Thomson problem on a Clifford torus in one, two, and three dimensions.
It uses the analytical expressions of the gradient and the Hessian of the Coulomb potential in Eq.~\ref{Eqn:totenergy}.
These expressions are reported in appendix \ref{GradientHessian}.
We note that our code could be easily generalized to higher dimensions but this is beyond the scope of our investigation.
The program uses the conjugated gradient algorithm to to minimize the Coulomb energy on the Clifford torus.
We use a total gradient-norm tolerance of $10^{-14}$ Hartree.
We verified that, at convergence, the eigenvalues of the Hessian of the Coulomb potential are all positive 
except for $n$ zero eigenvalues that correspond to the translational degrees of freedom.
To ensure that we find the global minimum and not a local minimum we have performed several calculations for a given Thomson problem starting from different initial positions of the electrons.
In practice, our code allows us to solve the Thomson problem numerically up to several thousands of electrons.

Our code can also be used to obtain information about structures that do not correspond to the global minimum of the potential energy surface by fixing the electrons to certain positions, e.g., those corresponding to a known crystal lattice.
After verifying that the gradient of $U$ vanishes, we can then learn from the Hessian whether this structure corresponds to a stationary point and, if this is the case, what is the nature of the stationary point (a minimum or a saddle point).
In the following, whenever the Hessian is analysed for a given configuration of electrons, it is implied that the gradient of $U$ vanishes for this configuration.
Finally, we note that the code with which all results in this work were obtained is freely available~\cite{Thomson_code}.

\section{Results}
\label{Sec:Results}
In this section we present the numerical solutions we have obtained for the Thomson problem on a Clifford torus.
We will focus here on two and three dimensions since the analytical result for one dimension is given in section \ref{Clifford1D}.
We note that we have verified that our numerical results in one dimension are consistent with the analytical results.
Although our method can be applied to the Thomson problem on a Clifford torus for any number of electrons, we focus here on a relatively large number of electrons (N $\sim$ 100) since we want to observe the lattice structures of Wigner crystals.
Finally, all the numerical solutions we present in this section, i.e., the positions of the electrons, can be found in Ref.~[\citenum{Thomson_code}].
\subsection{2-dimensional Clifford tori}
\begin{figure}
    \centering
    \includegraphics[scale=0.5]{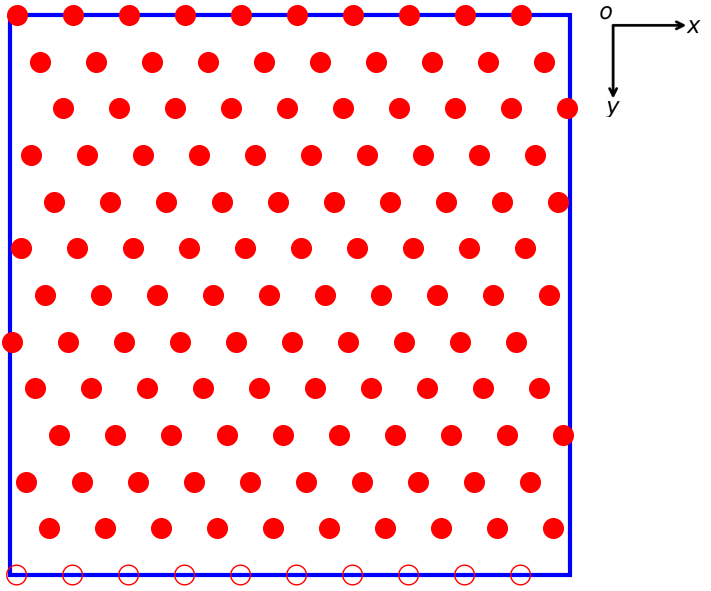}
    \caption{The solution to the Thomson problem for 120 electrons on a square Clifford torus ($r=1$). 
    Filled circles: the positions of the electrons; open circles: the equivalent positions of the electrons on the edge of the Clifford torus.
    }
    \label{FIG1}
\end{figure}

\begin{figure}
  \centering
  \includegraphics[scale=0.5]{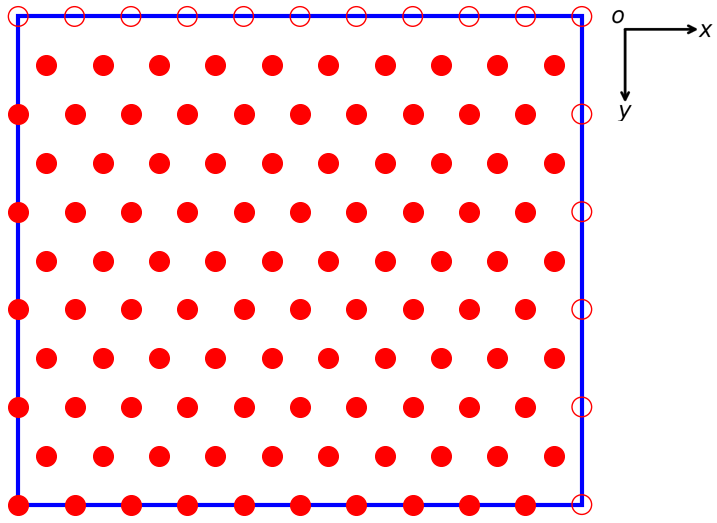}
  \caption{The solution to the Thomson problem for 100 electrons on a Clifford torus ($r=\sqrt{3}/2$). 
      Filled circles: the positions of the electrons; open circles: the equivalent positions of the electrons on the edge of the Clifford torus.
  The positions of the electrons correspond to a hexagonal latice.}
\label{FIG2}
\end{figure}

Although Eq.~\eqref{Eqn:totenergy} depends on the lengths of the two edges of the Clifford torus, $L_x$ and $L_y$, 
the relative positions corresponding to the solution of the Thomson problem only depend on the ratio $r=L_y/L_x$. Therefore, in the following we will only refer to $r$ and $N$ when discussing the solutions to the Thomson problem in 2D.
In Fig.~\ref{FIG1} we report the solution of the Thomson problem for $r=1$ and $N=120$.
Even though $r=1$ is not compatible with a hexagonal lattice, we clearly observe a hexagonal lattice with a small distortion.
The hexagonal lattice is commensurable with the Clifford torus when $r=\sqrt{3}/2$ and $N=4m^2$ with $m$ a positive integer.
In Fig.~\ref{FIG2} we report the solution to the Thomson problem on a Clifford torus with 100 electrons ($m=5$) and $r=\sqrt{3}/2$.
We now observe a perfect hexagonal lattice.
In general, we find that for $r=\sqrt{3}/2$ and $N=4m^2$ the solution to the Thomson problem is a hexagonal lattice for $m>1$.
Finally, we note that there are other combinations of $r$ and $N$ that are commensurable with the hexagonal lattice, e.g., $r=\sqrt{3}$ and $N=2m^2$.

With our approach we can also verify that the square lattice is not a solution to the Thomson problem on the Clifford torus by imposing a square lattice and calculating the corresponding Hessian.
A square lattice is commensurable with a Clifford torus Where $r=1$ and $N=m^2$ with $m$ a positive integer.
We have verified that the square lattice does not correspond to a minimum but to a saddle point, since the 
the corresponding Hessian has negative eigenvalues.
Our results are consistent with the prediction that the lattice structure of the two-dimensional Wigner crystal is the hexagonal lattice.
As mentioned before, this prediction is based on the comparison of the ground-state energy of the hexagonal lattice with other known lattices and, in particular, the square lattice.
In our approach the hexagonal lattice emerges naturally.
\subsection{3-dimensional Clifford tori}
\begin{figure}
\centering
\begin{subfigure}{.5\textwidth}
  \centering
  \includegraphics[scale=0.45]{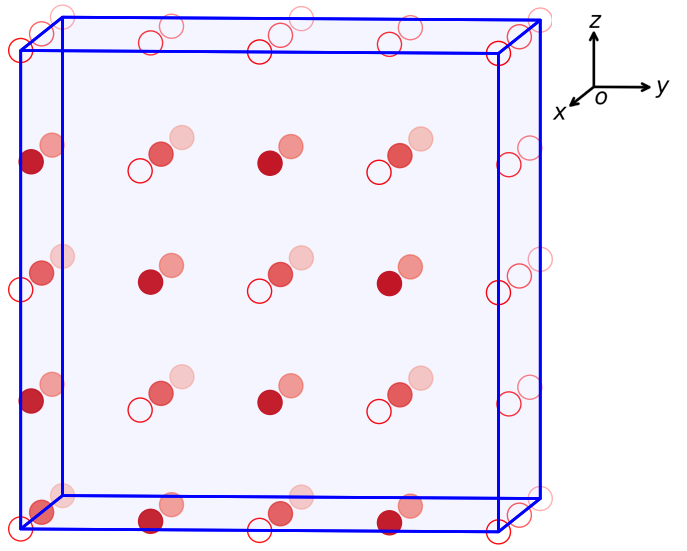}
  \label{fig:sub1}
\end{subfigure}%
\begin{subfigure}{.5\textwidth}
  \centering
  \includegraphics[scale=0.45]{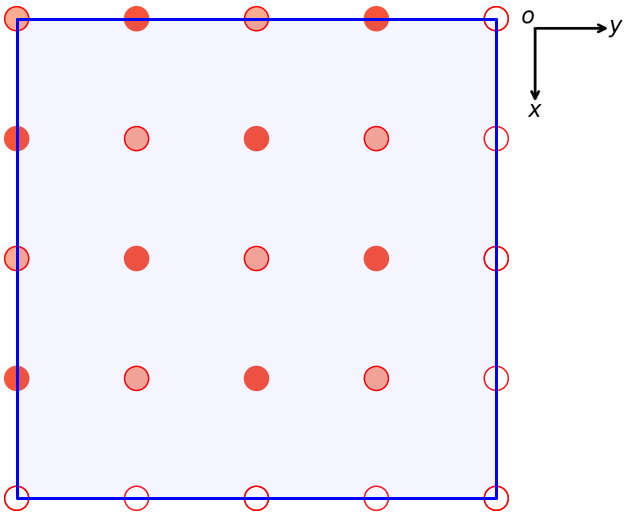}
  \label{fig:sub2}
\end{subfigure}
\caption{The solution to the Thomson problem for 32 electrons on a cubic Clifford torus. 
 Filled circles: the positions of the electrons (lighter shades represent electrons that are deeper); open circles : the equivalent positions of the electrons on the faces of the Clifford torus.
Left panel: front view;  right panel: top view. The positions of the electrons correspond to an fcc latice.}
\label{FIG3}
\end{figure}

In three dimensions the relative positions of the electrons corresponding to the solution of the Thomson problem on a Clifford torus depend on the number of electrons $N$ and on the ratios $L_y/L_x$ and $L_z/L_x$.
Here we will focus on cubic Clifford tori, i.e., $L_x = L_y = L_z$, with numbers of electrons that are commensurable either with the body-centred cubic (bcc) lattice, i.e., $N=2m^3$, or with the face-centred cubic (fcc) lattice, i.e., $N=4m^3$, with $m$ a positive integer. We note that $N$ can never be commensurable with both the bcc and fcc lattices.
We find that the fcc lattice is the solution to the Thomson problem for 4 ($m=1$), 32 ($m=2$), and 108 ($m=3$) electrons.
As an example, we report in Fig.~\ref{FIG3} the solution to the Thomson problem for a cubic Clifford torus with 32 electrons ($m=2$).
We have also verified that the fcc lattice is (at least) a local minimum for values of $4 \le m \le 10$ since the  corresponding Hessian matrices have no negative eigenvalues for these values of $m$.

\begin{figure}
\centering
\begin{subfigure}{.5\textwidth}
  \centering
  \includegraphics[scale=0.4]{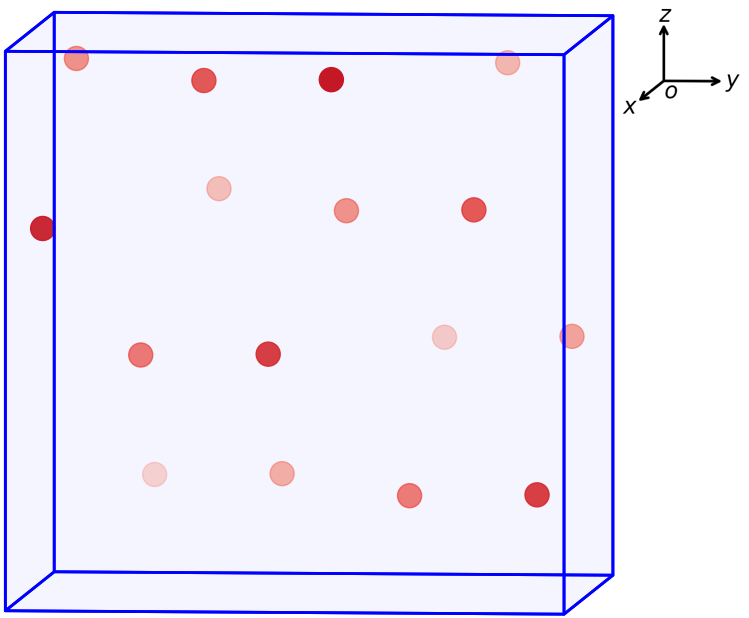}
  \label{fig:sub1}
\end{subfigure}%
\begin{subfigure}{.5\textwidth}
  \centering
  \includegraphics[scale=0.4]{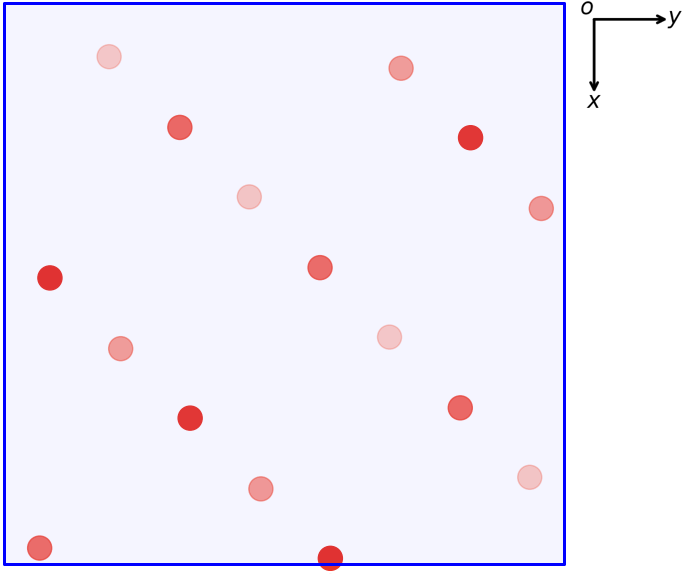}
  \label{fig:sub2}
\end{subfigure}
\caption{The solution to the Thomson problem for 16 electrons on a cubic Clifford torus. 
 Filled circles: the positions of the electrons (lighter shades represent electrons that are deeper);.
Left panel: front view;  right panel: top view. }
\label{FIG4}
\end{figure}

\begin{figure}
\centering
\begin{subfigure}{.5\textwidth}
  \centering
  \includegraphics[scale=0.45]{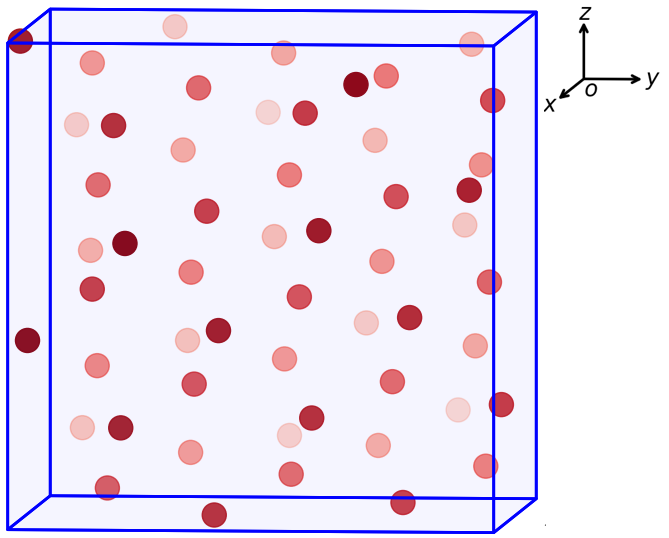}
  \label{fig:sub1}
\end{subfigure}%
\begin{subfigure}{.5\textwidth}
  \centering
  \includegraphics[scale=0.45]{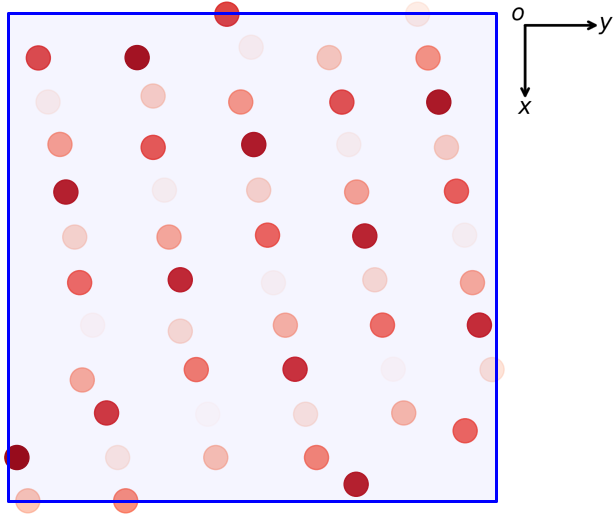}
  \label{fig:sub2}
\end{subfigure}
\caption{The solution to the Thomson problem for 54 electrons on a cubic Clifford torus. 
 Filled circles: the positions of the electrons (lighter shades represent electrons that are deeper).
Left panel: front view;  right panel: top view. }
\label{FIG5}
\end{figure}

\begin{figure}
\centering
\begin{subfigure}{.5\textwidth}
  \centering
  \includegraphics[scale=0.45]{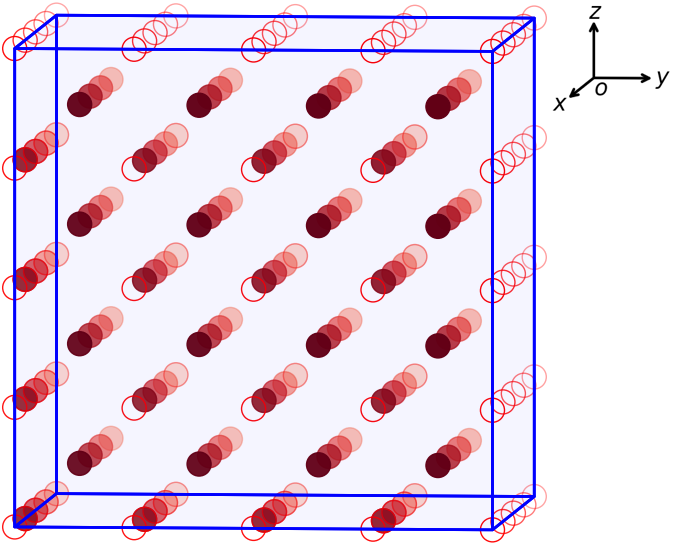}
  \label{fig:sub1}
\end{subfigure}%
\begin{subfigure}{.5\textwidth}
  \centering
  \includegraphics[scale=0.45]{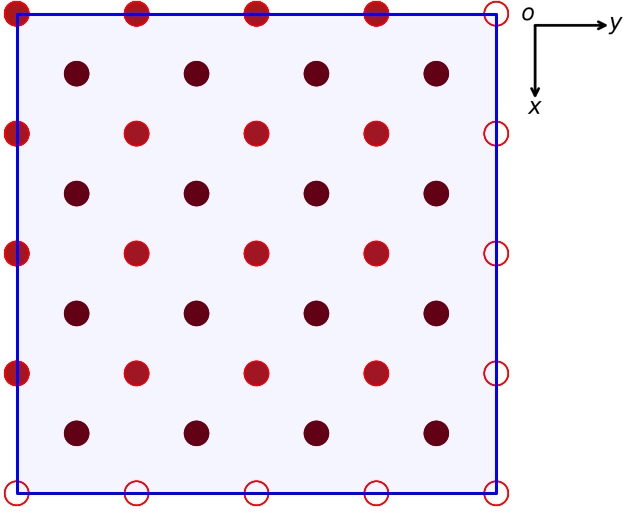}
  \label{fig:sub2}
\end{subfigure}
\caption{The solution to the Thomson problem for 128 electrons on a cubic Clifford torus. 
 Filled circles: the positions of the electrons (lighter shades represent electrons that are deeper); open circles: the equivalent positions of the electrons on the faces of the Clifford torus.
Left panel: front view;  right panel: top view. The positions of the electrons correspond to a bcc latice.}
\label{FIG6}
\end{figure}

Instead, for a number of electrons compatible with the bcc lattice the situation is more complicated.
With only 2 electrons $(m=1)$ the solution to the Thomson problem is the bcc lattice, with one electron on a vertex and the other in the center of the cube.
Instead for 16 electrons ($m=2$) the bcc lattice is not a minimum but a saddle point.
In Fig.~\ref{FIG4} we show the solution to the Thomson problem on a cubic Clifford torus with 16 electrons.
The electrons localise in planes that are parallel to the plane containing the diagonal of the cube.
In Fig.~\ref{FIG5} we show the solution to the Thomson problem on a cubic Clifford torus with 54 electrons ($m=3$).
In this case the bcc lattice corresponds to a local minimum but not to the global minimum.
For 128 electrons ($m=4$) the bcc lattice is again the solution to the Thomson problem.
We report that solution in Fig.~\ref{FIG6}; we clearly see that the electrons form a body-centred cubic (bcc) lattice.
Furthermore, we have verified that the bcc lattice is (at least) a local minimum for values of $5 \le m \le 13$ since the corresponding Hessian matrices have no negative eigenvalues for these values of $m$.
Finally, we have also verified that the simple cubic (sc) lattice is never a solutions to the Thomson problem on a cubic Clifford torus even when the number of electrons is commensurable with the sc lattice, i.e., $N=m^3$. Indeed, we find that these structures correspond to saddle points in the potential energy surface.

\section{The Thomson problem on a one-dimensional Clifford torus}
\label{Clifford1D}
In this section, we consider the analytical treatment of Thomson problem of $N$ electrons on a 1D Clifford torus of length $L$. 
Without loss of generality, we assume that the positions 
$x_i$ (modulus $L$) of the electrons $i$ are given in ascending order according to
\begin{equation}
    i < j \; \Rightarrow \; x_i < x_j \; \; \; ({\rm mod} \, L) .
\end{equation}
In 1D we can write the total energy $U$ as
\begin{equation}
    U = \frac{1}{2} \frac{\pi}{L} \sum_{i=1}^N \sum_{\substack{j=1 \\ j \neq i}}^N \sin\left[\frac{\pi}{L}|x_i-x_j|\right]^{-1}.
\end{equation}
It can be verified that this expression is equal to Eq.~\eqref{Eqn:totenergy} for values of $x_i$ (and $x_j$) on the Clifford torus to which the electrons are confined, i.e., $0 \le x_i \le L$.

We now show that the equally spaced arrangement of $N$ electrons on the 1D Clifford torus 
is a stationary point in the one-dimensional energy surface.
The gradient of the potential energy has the following components
\begin{equation}
    g_k = \frac{\partial U}{\partial x_k} = 
    \left(\frac{\pi}{L}\right)^2\sum_{\substack{j=1 \\ j\neq k}}^N
    \frac{\cos\left[\frac{\pi}{L}(x_k-x_j)\right]}{\sin^2\left[\frac{\pi}{L}|x_k-x_j|\right]} \text{sgn}(x_k - x_j).
\end{equation}
Each term of the sum in the above equation is an odd function of the argument $x_k-x_j$~.
Therefore, if the electrons are equally spaced, for each value $x_k-x_j$ there will be a symmetrical value $x_k - x_{j'} = -(x_k-x_j)$, and the two corresponding contribution cancel.
If the sum contains an odd number of terms there will be one contribution that will not be canceled.
However, this contribution corresponds to $x_k-x_j = L/2$ which vanishes by itself.
As a result, in the case of equally spaced electrons, each component $g_k$ vanishes, and this configuration necessarily corresponds to a stationary point.

We now demonstrate that this stationary point is a minimum. 
This means that all the eigenvalues of the Hessian are positive with the exception of one vanishing eigenvalue that 
corresponds to the translation of all the electrons.
The matrix elements $\mathcal{H}_{k,l}$ of the Hessian are given by
\begin{align}
\mathcal{H}_{k,k} &= \frac{\partial^2 U}{\partial x_k^2} = 
   \left(\frac{\pi}{L}\right)^3\sum_{\substack{j=1 \\ j\neq k}}^N
    \frac{1+\cos^2\left[\frac{\pi}{L}(x_k-x_j)\right]}{\sin^3\left[\frac{\pi}{L}|x_k-x_j|\right]},
\\
\mathcal{H}_{k,l} &= \frac{\partial^2 U}{\partial x_k \partial x_l} = 
   - \left(\frac{\pi}{L}\right)^3
    \frac{1+\cos^2\left[\frac{\pi}{L}(x_k-x_l)\right]}{\sin^3\left[\frac{\pi}{L}|x_k-x_l|\right]} \quad\quad (k\neq l).
\end{align}
We see that the Hessian is real and symmetric. All its diagonal elements are positive while all off-diagonal elements are negative.
Since all electrons are equivalent when they are equally spaced on a Clifford torus, all diagonal elements are equal to the same value.
Moreover, all off-diagonal elements with the same value for $|k-l|$ have identical values.
Therefore, we can write
\begin{equation}
\mathcal{H}_{k,l} = H_{|k-l|} .
\end{equation}
Finally, because of the translational symmetry of the Clifford torus we have the identity
\begin{equation}
H_j = H_{N-j}.
\end{equation}
By combining the above results it follows that the Hessian can be rewritten as a symmetric circulant matrix according to
\begin{equation}
\mathcal{H} = 
\left\|
\begin{array}{ccccccccc}
H_0   & H_1   & H_2  & H_3 & ... & H_{N-2} & H_{N-1}  \\
H_{N-1}   & H_0   & H_1  & H_2 & ... & H_{N-3} & H_{N-2}  \\
H_{N-2}   & H_{N-1}   & H_0  & H_1 & ... & H_{N-4} & H_{N-3}  \\
...   & ...   & ...  & ... & ... & ... & ...  \\
H_2   & H_3   & H_4  & H_5 & ... & H_0 & H_1  \\
H_1   & H_2   & H_3  & H_4 & ... & H_{N-1} & H_0  \\
\end{array}
\right\|
=
\left\|
\begin{array}{ccccccccc}
H_0   & H_1   & H_2  & H_3 & ... & H_2 & H_1  \\
H_1   & H_0   & H_1  & H_2 & ... & H_3 & H_2  \\
H_2   & H_1   & H_0  & H_1 & ... & H_4 & H_3  \\
...   & ...   & ...  & ... & ... & ... & ...  \\
H_2   & H_3   & H_4  & H_5 & ... & H_0 & H_1  \\
H_1   & H_2   & H_3  & H_4 & ... & H_1 & H_0  \\
\end{array}
\right\|.
\end{equation}
in which 
\begin{align}
    H_0(N) &= \frac{\pi^3}{d^3 N^3}\sum_{j=1}^{N-1} \frac{1 + \cos^2 \left(\frac{\pi j}{N}\right)}
    {\sin^3 \left|\frac{\pi j}{N}\right|},
    \\
    H_j (N) &= - \frac{\pi^3}{d^3 N^3}\frac{1 + \cos^2 \left(\frac{\pi j}{N}\right)}{\sin^3 \left| \frac{\pi j}{N}\right|} \quad  (j\neq 0),
\end{align}
where we used that $L = Nd$ and $(x_k - x_l) = (k-l)d$ with $d$ the nearest-neighbor distance of the electrons.
From the above equations it can be easily verified that the following sum rule holds,
\begin{equation}
H_0 = - \sum_{j=1}^{N-1} H_j.
\label{Eqn:sumrule}
\end{equation}

Since $\mathcal{H}$ is a symmetric circulant matrix, the normalized eigenvectors are given by
\begin{equation}
\phi_k(N)
= \frac{1}{\sqrt{N}}
\left\|
\begin{array}{ccccccccc}
1   \\
e^{i k\theta} \\
e^{2i k\theta} \\
...   \\
e^{i(N-2) k\theta} \\
e^{i(N-1) k\theta} \\
\end{array}
\right\|
= \frac{1}{\sqrt{N}} \;
\left\|
\begin{array}{ccccccccc}
1   \\
e^{i k\theta} \\
e^{2i k\theta} \\
...   \\
e^{-2i k\theta} \\
e^{-i k\theta} \\
\end{array}
\right\|,
\end{equation}
where $k=0,...,N-1$, and $\theta(N) = 2\pi/N$.
It follows that the corresponding eigenvalues of $\mathcal{H}$ are given by
\begin{equation}
    \epsilon_k (N) = \sum_{j=0}^{N-1} \cos(k j \theta) H_j,
\end{equation}
where we used Euler's formula.
Let us first focus on $k=0$.
The corresponding eigenvalue $\epsilon_0$ is given by
\begin{equation}
    \epsilon_0 (N) = \sum_{j=0}^{N-1} H_j = 0,
    \label{epsilon_0}
\end{equation}
which follows from the sum rule in Eq.~\eqref{Eqn:sumrule}.
The corresponding eigenvector $\phi_0(N)$ is given by
\begin{equation}
\phi_0(N) = \frac{1}{\sqrt{N}}
\left\|
\begin{array}{ccccccccc}
1   \\
1 \\
1 \\
...   \\
1 \\
1 \\
\end{array}
\right\|,
\end{equation}
which describes a collective displacement of all the electrons in the same direction and by the same amount.
This is consistent with $\epsilon_0=0$ since a collective displacement does not modify the total energy of the system.
Let us now study the eigenvalues for $k>0$.
Since all $H_j$ are negative for $j>0$ we have the following inequality,
\begin{equation}
 - \sum_{j=1}^{N-1} \cos(k j \theta) H_j < - \sum_{j=1}^{N-1} H_j = H_0,
    \label{epsilon_k}
\end{equation}
where we used once more the sum rule in Eq.~\eqref{Eqn:sumrule},
it follows that 
\begin{equation}
    \epsilon_k (N) > 0 \quad (\forall k>0) .
\end{equation}
Therefore, the $\epsilon_k$~, with $k>0$, are all strictly positive.
This concludes the proof that a one-dimensional lattice of equally spaced electrons on the Clifford torus is indeed a local minimum.

\begin{figure}
%    \centering
    \includegraphics[scale=1.2]{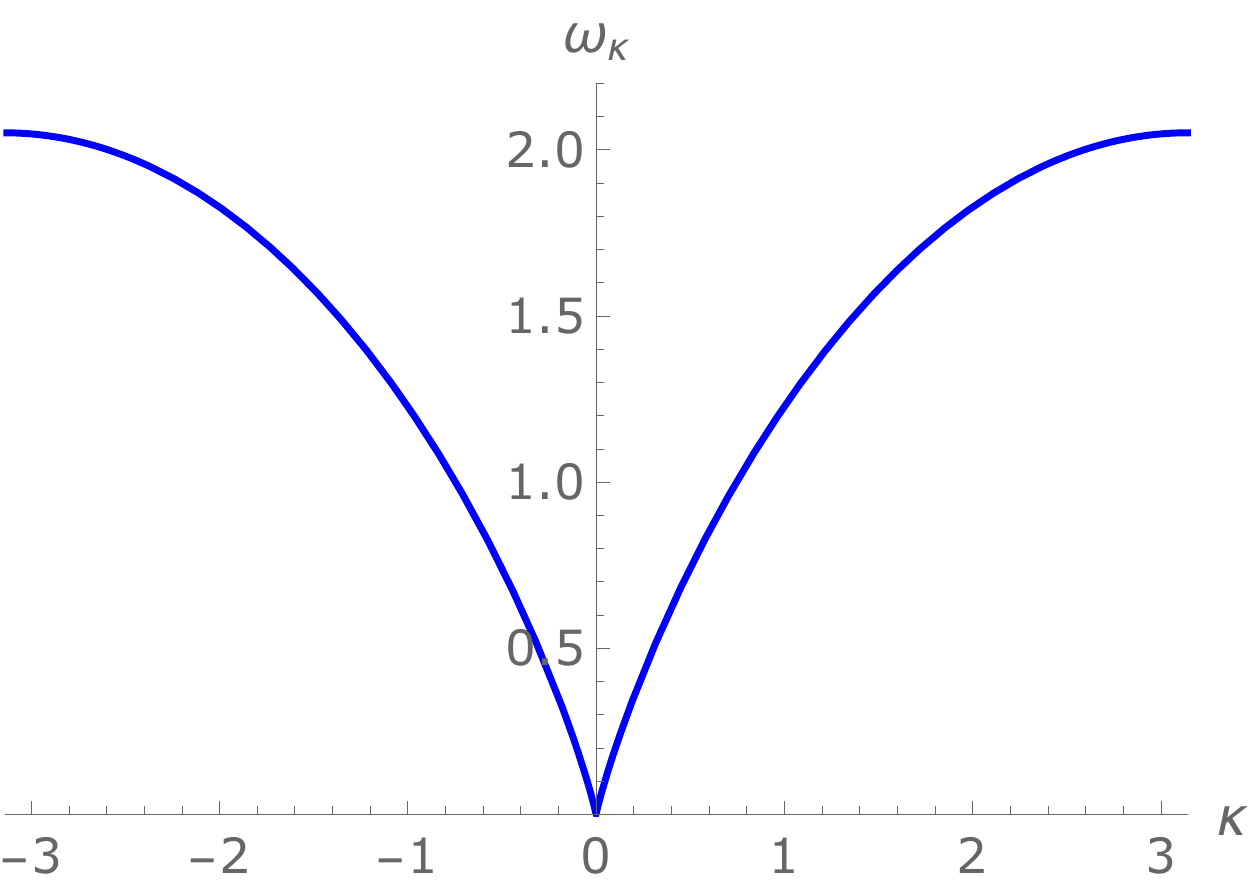}
    \caption{The angular frequency $\omega_{\kappa}$ of a one-dimensional Wigner crystal as a function of the lattice momentum $\kappa$.
    To put into evidence its behaviour around $\kappa=0$, we show $\omega_{\kappa}$ for $ \kappa \in [-\pi, \pi$].
    }
    \label{Fig:dispersion}
\end{figure}

Finally, it is interesting to evaluate what happens in the thermodynamic limit $N\rightarrow\infty$.
The eigenvalues become
\begin{align}
    \lim_{N\rightarrow\infty}\epsilon_k (N) & = 
    \lim_{N\rightarrow\infty} H_0 (N) + \lim_{N\rightarrow\infty} \sum_{j=1}^{N-1} \cos(k j \theta) H_j (N)
    \\ &=
    \frac{2}{d^3} \sum_{j=1}^{\infty} \frac{1- \cos(j \kappa)}{j^3},
\end{align}
where $\kappa= 2\pi k / N$ is the crystal momentum, with $0\le \kappa < 2\pi$, and we used
\begin{align}
    \lim_{N\rightarrow\infty} H_0(N) &= \sum_{j=1}^\infty \frac{2}{d^3 j^3}
    \\
    \lim_{N\rightarrow\infty} H_j(N) &=- \frac{2}{d^3 j^3}.
\end{align}
The summation in the above equation can be performed analytically and the result is
\begin{equation}
    \epsilon_{\kappa} = 
    \frac{1}{d^3} \left(2\zeta(3) -\left({\rm Li}_3(e^{i\kappa}) + {\rm Li}_3(e^{-i\kappa}) \right) \right) ,
\end{equation}
where $\zeta(z)$ is the Riemann zeta function and ${\rm Li}_3(z)$ is the trilogarithm function.
It can be easily verified that $\epsilon_{\kappa}$ is an even function of $\kappa$.
In Fig.~\ref{Fig:dispersion} we report the angular frequencies $\omega_{\kappa} = \sqrt{\epsilon_{\kappa}}$ as a function of the crystal momentum $\kappa$ corresponding to $d=1$ Bohr.
To put into evidence its behaviour around $\kappa=0$, we show $\omega_{\kappa}$ for $ \kappa \in [-\pi, \pi$].
At first sight, the curve has a broad similarity with the typical behaviour of $\omega_\kappa$
as a function of $\kappa$ for the case of harmonic crystals, as can be found in many textbooks. 
In that case the derivative  $\frac{d \omega_\kappa }{ d \kappa}$ tends to a finite constant value when $\kappa\rightarrow0$. 
It corresponds  to the speed of sound at low frequencies.
Instead, the curve in Fig.~\ref{Fig:dispersion} for the one-dimensional Wigner crystal, has an infinite slope because $\frac{d \omega_\kappa }{ d \kappa}$ diverges logarithmically when $\kappa\rightarrow0$
as can be seen from the Taylor expansion of $\epsilon_{\kappa}$ around $\kappa=0$ given by
\begin{equation}
\epsilon_{\kappa} = \frac{1}{d^3} \left[\left(\frac32- \ln|\kappa|\right)\kappa^2 + \frac{\kappa^4}{144}
+ \frac{\kappa^6}{43 \, 200} + \frac{\kappa^8}{5 \, 080 \, 320}  + \mathcal{O}(\kappa^{10}) \right].
\end{equation}
Strictly speaking, it corresponds to an infinite speed of sound in the limit of low frequencies.
We note that, because of the logarithmic nature of the singularity, it cannot be observed in Fig.~\ref{Fig:dispersion}.
\section{Conclusions}
\label{Sec:Conclusions}
In this work, we presented a study of the Thomson problem, in which electrons interact through a repulsive Coulomb interaction, on $n$-dimensional Clifford Tori ${\cal T}^n$, with $n$ = 1, 2, 3,
Similarly to what is usually done for the Thomson problem on the surface of the ordinary sphere, ${\cal S}^2$, the distance that was used in the Coulomb law is the Euclidean distance in the embedding space of the torus, which in this case is the $n$-dimensional complex space $\mathbb{C}^n$.
We applied our approach to the search of the minimum-energy configurations of Wigner crystals.
Our approach permits to simulate a fragment of an infinite Wigner crystal, while avoiding the border effects that are necessarily associated to a finite-cluster calculation within open-boundary conditions.
We performed a series of numerical calculations in which we found the minimum Coulomb energy of the system starting from random initial geometries.
We showed that, for a sufficiently large number of electrons, the global minima correspond to the linear lattice (1D), the hexagonal lattice (2D) and the body-centred cubic lattice (3D).
We cannot exclude, in principle, that other regular structures having lower energies can exist, for instance, by  increasing even further the number of electrons.
However, our calculations are in agreement with the existing results in the literature on the prediction of the lattice structures of Wigner crystals.

In future works, we plan to generalize the Thomson problem on Clifford tori to different inter-particle potentials.
In particular, we will study the behavior of clusters bounded by harmonic potentials.
In the limit of systems having a small size with respect to the torus dimension, in fact, the physics of the system should converge to that of the same systems in the ordinary infinite space. 
In a more general perspective, it will be possible to use completely general distance-dependent particle-particle interactions and replace the ordinary distance in the infinite open space by the Euclidean distance in a toroidal supercell. In this a way, one could perform molecular mechanics simulations on a set of particles, or molecules, in a periodic context.
Finally, we will extend our approach to a full quantum treatment of the electrons by extending the approach described in Ref.~[\citenum{escobar2021wigner}] to more than 2 electrons.
\section{Acknowledgements}

We thank the French ``Agence Nationale de la Recherche (ANR)'' for financial
support (Grant Agreements No. ANR-19-CE30-0011 and ANR-22-CE29-0001). This work
has been (partially) supported through the EUR grant NanoX n$^\circ$
ANR-17-EURE-0009 in the framework of the ``Programme des Investissements
d'Avenir''. 
\appendix
\section{The gradient and Hessian of the Coulomb energy}
\label{GradientHessian}
The gradient of the Coulomb energy given in Eq.~\eqref{Eqn:totenergy} has the following components
\begin{equation}
    \partial_{x_{\alpha,i}}  U = - \frac{1}{2} \left(\frac{\pi}{L_{\alpha}}\right)  
    \sum_{\substack{k=1 \\ k\neq i}}^N
    \left[ \sum_{\beta=1}^n \left(\frac{L_{\beta}}{\pi}\right)^2 \sin^2 \left( \frac{\pi}{L_{\beta}} (x_{\beta,i}-x_{\beta,k})\right)  \right]^{-3/2} \sin\left(\frac{2\pi}{L_{\alpha}}(x_{\alpha,i}-x_{\alpha,k})\right).
\end{equation}
The Hessian of the Coulomb energy has the following components,
\begin{align}
    \partial^2_{x_{\alpha,i} x_{\alpha,i}} U & =  \frac{3}{2} \frac{\pi^2}{L_{\alpha}^2}  \sum_{\substack{k=1 \\ k\neq i}}^N \left[\sum_{\beta=1}^n \left(\frac{L_{\beta}}{\pi}\right)^2 \sin^2 \left( \frac{\pi}{L_{\beta}} (x_{\beta,i}-x_{\beta,k})\right) \right]^{-5/2} \sin^2\left(\frac{2\pi}{L_{\alpha}}(x_{\alpha,i}-x_{\alpha,k})\right)
    \nonumber \\ &-
    \sum_{\substack{k=1 \\ k\neq i}}^N \left[ \sum_{\beta=1}^n \left(\frac{L_{\beta}}{\pi}\right)^2 \sin^2 \left(\frac{\pi}{L_{\beta}} (x_{\beta,i}-x_{\beta,k})\right) \right]^{-3/2}  \cos\left(\frac{2\pi}{L_{\alpha}}(x_{\alpha,i}-x_{\alpha,k})\right),
    \\
    \partial^2_{x_{\alpha,i} x_{\alpha,j}} U &=  -\frac{3}{2} \left(\frac{\pi}{L_{\alpha}}\right)^2 \sum_{\substack{k=1 \\ k\neq i}}^N \left[\sum_{\beta=1}^n \left(\frac{L_{\beta}}{\pi}\right)^2 \sin^2 \left( \frac{\pi}{L_{\beta}} (x_{\beta,i}-x_{\beta,j})\right) \right]^{-5/2} \sin^2\left(\frac{2\pi}{L_{\alpha}}(x_{\alpha,i}-x_{\alpha,j})\right)
    \nonumber  
    \nonumber \\& +
       \sum_{\substack{k=1 \\ k\neq i}}^N \left[ \sum_{\beta=1}^n \left(\frac{L_{\beta}}{\pi}\right)^2 \sin^2 \left( \frac{\pi}{L_{\beta}} (x_{\beta,i}-x_{\beta,j})\right) \right]^{-3/2} \cos\left(\frac{2\pi}{L_{\alpha}}(x_{\alpha,i}-x_{\alpha,j})\right), \quad (i\neq j)
    \\
    \partial^2_{x_{\alpha,i} x_{\beta,j}} U &=  \frac{3}{4} \frac{\pi^2}{L_{\alpha}L_{\beta}} \left[ \sum_{\gamma=1}^n \left(\frac{L_{\gamma}}{\pi}\right)^2 \sin^2 \left(\frac{\pi}{L_{\gamma}} (x_{\gamma,i}-x_{\gamma,j})\right) \right]^{-5/2} \nonumber
    \\ & \times \sin\left(\frac{2\pi}{L_{\alpha}}(x_{\alpha,i}-x_{\alpha,j})\right) \sin\left(\frac{2\pi}{L_{\beta}}(x_{\beta,i}-x_{\beta,j})\right) \quad (i\neq j, \alpha \neq \beta).
\end{align}
We note that the Hessian matrix can be diagonalized to gain information about the nature of the various stationary points on the energy surface from the eigenvalues and eigenvectors.
\providecommand{\latin}[1]{#1}
\makeatletter
\providecommand{\doi}
  {\begingroup\let\do\@makeother\dospecials
  \catcode`\{=1 \catcode`\}=2 \doi@aux}
\providecommand{\doi@aux}[1]{\endgroup\texttt{#1}}
\makeatother
\providecommand*\mcitethebibliography{\thebibliography}
\csname @ifundefined\endcsname{endmcitethebibliography}
  {\let\endmcitethebibliography\endthebibliography}{}

%
%\bibliography{main.bib}
%
\end{document}